\documentstyle[aps,preprint,tighten]{revtex}
\def\e{{\rm e}}
\def\i{{\rm i}} 
\def\d{{\rm d}}
\begin{document}
\title{Strong Electron Tunneling through Mesoscopic Metallic Grains}
\author{D.S. Golubev$^a$, J\"urgen K\"onig$^b$, Herbert Schoeller$^b$, 
Gerd Sch\"on$^b$, and  A.D. Zaikin$^{a,b}$}
\date{\today}

\address{$^a$ I.E.Tamm Department of Theoretical Physics,\\
P.N.Lebedev Physics Institute, Leninskii pr. 53, 117924
Moscow, Russia \\
$^b$ Institut f\"ur Theoretische Festk\"orperphysik,
Universit\"at Karlsruhe, 76128 Karlsruhe, Germany}
\maketitle
\input psfig
\maketitle
\begin{abstract}
We describe electron transport through small metallic grains with 
Coulomb blockade effects beyond the perturbative regime. For this purpose
we study the real-time evolution of the reduced density matrix of the system.
In the first part of the paper we present a diagrammatic expansion 
for not too high junction conductance, 
$h/4\pi^2e^2 R_{\rm t} \lesssim 1$, in a basis of charge states.
Quantum fluctuations renormalize system parameters and lead to 
finite lifetime 
broadening in the gate-voltage dependent differential conductance. 
We derive  analytic results for the spectral density and the conductance in
the limit where only two charge states play a role.
In the second part of the paper we consider junctions with large conductance, 
$h/4e^2 R_{\rm t} \gtrsim 1$. 
In this case contributions from all charge states, which broaden and overlap,
become important. We analyze the problem in a quasiclassical approximation.
The two complementary approaches cover the essential features of 
electron tunneling for all parameters.

\end{abstract}

\section{Introduction}

Electron transport through mesoscopic grains is strongly 
influenced by the large charging energy, $E_{\rm C}= e^2/2C$,
associated with the low capacitance $C$ of the 
system~\cite{AL,SZ,gra-dev,Karlsruhe}. 
An interesting example is the ``single-electron transistor'' where a small 
metallic island is coupled via tunnel junctions to leads and 
via a capacitor to a gate voltage. At low temperatures, $T \ll E_{\rm C}$,
a variety of single-electron phenomena
have been observed in this system, including the Coulomb blockade and 
oscillations of the conductance as a function of a gate voltage.
If the dimensionless conductance of the tunnel junctions 
between the island and the lead electrodes,
\begin{equation}
	\alpha_{\rm t} \equiv \frac{R_{\rm K}}{4\pi^2R_{\rm t}} 
	= \frac{h}{4\pi^2 e^2 R_{\rm t}} \, ,
\label{alpha}
\end{equation}
is small, on a scale defined by the quantum resistance 
$R_{\rm K}\simeq 25.8$~k$\Omega $, the charge of each island is a well-defined 
variable.  
In the limit $\alpha_{\rm t}\ll 1$, the sequential single-electron tunneling 
can be studied in perturbation theory~\cite{AL,gra-dev}; and descriptions 
based on a master equation or equivalent simulations of the stochastic 
dynamics are sufficient to account for the dominant features observed in 
single-electron devices.

Recent experiments beyond the perturbative regime show deviations from
the classical description, e.g. a broadening of the 
conductance peaks much larger than temperature \cite{est,Kuz}.
This indicates that, in general, quantum fluctuations and higher-order 
coherent processes should be considered.
Even in the limit of weak tunneling, $\alpha_{\rm t} < 1$, 
nontrivial features
appear in the vicinity of the Coulomb blockade threshold, when two 
charge states become nearly degenerate and perturbation theory fails. 
Several theoretical papers 
\cite{M,GZ94,schoell-schoen,Grabert1,Falci,koe-schoell-schoen1,koe-schoell}
dealt with the problem of higher-order 
processes, exploiting the physical picture of electron tunneling via 
discrete charge states.
This includes ``inelastic cotunneling''~\cite{Av-Naz,koe-schoell}, 
where in a second-order process in the parameter 
$\alpha_{\rm t}$ electrons tunnel via a virtual state of the island. 
An extension of this process, which gains importance near resonances, is 
``inelastic resonant tunneling''~\cite{schoell-schoen,koe-schoell-schoen1}, 
a process where electrons tunnel an arbitrary number of
times between the reservoirs and the islands. 
The term ``inelastic'' indicates that with overwhelming 
probability different electron states are involved in the 
different steps of the higher order processes. The description can been
 extended to describe strong tunneling through single level 
quantum dots~\cite{koe-schoell-schoen2}.

If the conductance of tunnel junctions is not small, 
$\alpha_{\rm t} \gtrsim 1$, the physical picture changes.
In this case the inverse lifetime $\Gamma = 1/R_{\rm t} C$ and, hence, the 
broadening of the excited charge states due to 
quantum fluctuations exceed the 
typical level spacing of excited island states, 
$\hbar \Gamma \gtrsim E_{\rm C}$.
Thus charge levels overlap and the concept of tunneling via discrete 
charge states becomes ill-defined, raising the question  whether charging 
effects survive under such conditions or whether they are washed out 
completely by strong quantum fluctuations. 
In Refs.~\onlinecite{Guinea,PZ88,SZ,PZ91,ZP93,Falci}
 it was demonstrated that at sufficiently low temperatures
 even for large values of $\alpha_{\rm t}$ quantum 
fluctuations of the charge {\it do not} destroy Coulomb 
blockade of tunneling, but they lead to a strong 
renormalization of the effective junction capacitance,
$ C_{\rm eff} \propto C \exp (2\pi^2\alpha_{\rm t})$.
The exponential dependence on $\alpha_{\rm t}$ had been derived 
independently by 
renormalization group arguments~\cite{Guinea,Falci}, instanton
techniques~\cite{PZ91}, and  Monte Carlo studies~\cite{Falci,WEG}.
One important consequence of the strong capacitance renormalization with 
increasing $\alpha_{\rm t} \gtrsim 1$ is the exponential reduction  of the 
temperature limit below which  charging effects can be observed.

This article is devoted to the calculation of the conductance of a SET 
transistor beyond perturbation theory in $\alpha_{\rm t}$, in a range of 
parameters which is accessible to experiments.
The island contains a large number of electrons which are coupled strongly by 
Coulomb interactions. 
We, therefore, cannot proceed with ordinary perturbation theory.
Rather, we reformulate the quantum mechanical many-body problem of these 
electrons in a real-time path-integral representation.
In order to handle the Coulomb interaction we perform a Hubbard-Stratonovich
transformation which introduces a phase as a collective variable.
We trace out all microscopic degrees of freedom and arrive at an effective
action of the system~\cite{Eckern,SZ}, similar in structure to that known from 
the studies of Ohmic dissipation in quantum mechanics~\cite{cal-leg}.
This procedure is addressed in Section 2.

After a change from the phase to a charge representation we are able to 
perform for $\alpha_{\rm t} \lesssim 1$ a diagrammatic expansion of the time 
evolution of the reduced density matrix.
In a charge representation we can identify sequential, co- and resonant 
tunneling processes with certain classes of diagrams.
A restriction to two charge states allows us to evaluate the spectral function 
and the conductance of the system analytically. 
The results will be presented in Section 3.
At higher temperatures more charge states play a role, which in general 
requires a numerical study of the diagrammatic expansion.

In the opposite limit of strong tunneling, $\alpha_{\rm t} \gtrsim 1$,
many charge states play a role, and a formulation in terms of the 
phase, which is canonically conjugated to the charge, is more convenient. 
This limit is discussed in Section 4. We analyze quantum dynamics of the 
phase variable in a semiclassical (saddle-point)
approximation and obtain an expression for the system conductance valid at 
not too low temperatures $T \gtrsim e^2/2C_{\rm eff}$. 
The exponential renormalization of the effective capacitance for strong 
tunneling widens this temperature range substantially.
The two approaches cover the essential features of electron tunneling
for all parameters.

In Section 5 we review  briefly some results obtained earlier within
different imaginary time techniques, e.g.
renormalization group and instanton
methods, and compare these results with those of our real time analysis.

\section{Formulation of the Problem}

We consider a metallic island coupled by two tunnel junctions (L,R) to two 
leads and capacitively to an external gate voltage $V_{\rm g}$.
An applied transport voltage $V=V_{\rm L}-V_{\rm R}$ drives a current.
A microscopic description of this single-electron 
transistor is based on the  Hamiltonian, 
$H=H_{\rm L}+H_{\rm R}+H_{\rm I}+H_{\rm ch}+H_{\rm t,L}+H_{\rm t,R}$. 
Here $H_{\rm r}= \sum\limits_{k\sigma}\epsilon_{k\sigma {\rm r}}
a^{\dagger}_{k\sigma {\rm r}}a_{k\sigma {\rm r}}$ 
describes noninteracting electrons in the left and right lead, 
r= L,R, and
$H_{\rm I}=\sum\limits_{q\sigma}\epsilon_{q\sigma}
c^{\dagger}_{q\sigma}c_{q\sigma}$ models the island states.
The Coulomb interaction is accounted for in a capacitance model
\begin{equation}
	H_{\rm ch}= E_{\rm C} \left( \sum\limits_{q\sigma}
	c^{\dagger}_{q\sigma}c_{q\sigma} -n_{\rm g} \right) ^2 \; .
\end{equation}	
The energy scale $E_{\rm C} \equiv e^2/(2C)$ of the transistor 
depends on the total island capacitance,
$C = C_{\rm L} + C_{\rm R} + C_{\rm g}$,
determined by the left and right tunnel junction and the gate
capacitance. 
The charging energy can be tuned continuously by the ``gate charge'' 
\begin{equation}
	Q_{\rm g}\equiv -en_{\rm g}=-(C_{\rm L}V_{\rm L}+C_{\rm R}V_{\rm R}
	                             +C_{\rm g}V_{\rm g}) \; .
\label{Qg}
\end{equation} 
The tunneling Hamiltonian 
$H_{\rm t,r}=\sum\limits_{kq \sigma}\left( T^{\sigma {\rm r}} 
a^{\dagger}_{k \sigma {\rm r}} c_{q \sigma}+{\rm h.c.}\right)$ describes 
tunneling between the island and the left and right leads.
The matrix elements are related to the tunnel conductances by 
$R_{\rm r}^{-1} = (e^2 / h) \sum\limits_\sigma N_{\rm r}^\sigma(0) 
N_{\rm I}^\sigma(0) |T^{\sigma {\rm r}}|^2$, 
where $N(0)$ denotes the densities of 
states of the island and the leads, respectively. 
In the following we will consider ``wide'' metallic junctions with $N \gg 1$ 
transverse channels.
Extending the spin summation they can be labeled by the index  
$\sigma = 1, ... N$.
In the following we will put $\hbar = 1$ (except when it enters the quantum of 
resistance).

Our aim is to study the time-evolution of the density matrix. 
We shortly sketch the main steps of the derivation of this description:\\
-- The time evolution of the density matrix introduces two propagators, a 
forward and backward propagator, which get coupled when we trace out
electron degrees of freedom of the reservoirs. 
The procedure is known from the work of Caldeira and Leggett~\cite{cal-leg} 
who, generalizing earlier work of Feynman and Vernon, 
studied the influence of Ohmic dissipation on a quantum system. 
Similarly the influence on electron tunneling was described in 
Refs.~\onlinecite{Eckern,SZ}. 
Here, we generalize the later work from a single tunnel junction to the 
transistor.\\
-- In order to describe the Coulomb interaction between electrons we introduce 
via a Hubbard-Stratonovich transformation the electric potential of the island 
$V(t)$ as a macroscopic field.
The interaction between electrons is replaced in this way by an interaction 
with the collective variable.\\
-- We treat the leads  as well as the electrons in the island as large 
equilibrium reservoirs. 
The electrochemical potentials of the reservoirs are fixed,
$\mu_{\rm r}=-eV_{\rm r}$ for r = L,R.
The only fluctuating field is voltage of the island $V(t)$.
The definition $eV(t)\equiv -\dot{\varphi}(t)$ relates $V(t)$ to a phase 
$\varphi (t)$.
Its quantum mechanical conjugate is the number of excess electrons $n(t)$ on 
the island. 
As a consequence of the procedure outlined so far, the macroscopic field 
$n(t)$ is independent of the microscopic degrees of freedom described by 
$c_{q\sigma}$ and $c^{\dagger}_{q\sigma}$.
At this stage, the electronic degrees of freedom can be traced out.\\
-- The time evolution of the reduced density matrix 
$\rho(t;\varphi_1,\varphi_2)$, which depends only on the phase variable 
$\varphi$, can thus be expressed by a double path integral over the phases 
corresponding to the forward and backward propagators $\varphi_j$ ($j=1,2$)
\begin{equation}
	\rho_{\rm c}(t_{\rm f};\varphi_{\rm 1f},\varphi_{\rm 2f})= 
	\int \limits_{-\infty}^{\infty} d\varphi_{\rm 1i} 
	\int \limits_{-\infty}^{\infty} d\varphi_{\rm 2i}
	\int\limits_{\varphi_{\rm 1i}}^{\varphi_{\rm 1f}}
	{\cal D}[\varphi_1(t)] 
	\int\limits_{\varphi_{\rm 2i}}^{\varphi_{\rm 2f}}
	{\cal D}[\varphi_2(t)]
	\exp\left(iS[\varphi_1(t),\varphi_2(t)]\right)
	\rho_{\rm c}(t_{\rm i};\varphi_{\rm 1i},\varphi_{\rm 2i}) \, .
\label{continuous}
\end{equation}
-- The form (\ref{continuous}) describes the situation where charges can take 
any continuous value and the phase is an extended variable. 
However, in our physical system the charge on the island is quantized in units 
of the electron charge $e$. 
In this case the phase variable is compact (i.e., the states $\varphi $ and 
$\varphi +2\pi $ are equivalent), and we rewrite (\ref{continuous}), 
introducing integer winding numbers $m_1, m_2 = 0, \pm 1, \pm 2, \dots$,
\begin{eqnarray}
	\rho_{\rm d}(t_{\rm f};\varphi_{\rm 1f},\varphi_{\rm 2f})= 
	\sum\limits_{m_1,m_2}
	\int \limits_{-\infty}^{\infty} d\varphi_{\rm 1i} 
	\int \limits_{-\infty}^{\infty} d\varphi_{\rm 2i}
	\int\limits_{\varphi_{\rm 1i}}^{\varphi_{\rm 1f}+2\pi m_1}
	{\cal D}[\varphi_1(t)]
	\int\limits_{\varphi_{\rm 2i}}^{\varphi_{\rm 2f}+2\pi m_2}
	{\cal D}[\varphi_2(t)]
	\nonumber \\
	\exp\left(iS[\varphi_1(t),\varphi_2(t)]\right)
	\rho_{\rm d}(t_{\rm i};\varphi_{\rm 1i},\varphi_{\rm 2i}) \, .
	\label{discrete}
\end{eqnarray}

The two integrations can be combined to a single integral along the Keldysh 
contour, which runs forward and backward between 
$t_{\rm i}$ and $t_{\rm f}$ along the real-time axis. 
As a result the reduced propagator $\Pi$ is written as a 
single path integral along this contour
\begin{equation}
	\Pi=\mbox{tr}\left[\rho_0 \,
	T_{\rm K} \exp \left(-i\int_{\rm K} \d t \, H(t) 
	\right)\right]=
	\int{\cal D}[\varphi(t)]\exp\left(iS[\varphi(t)]\right).
\end{equation}
Here  the collective variable $\varphi(t)$ and  the time integral 
are defined on the Keldysh contour K, and the time-ordering 
operator $T_{\rm K}$ orders the following operators accordingly. 

The effective action entering the propagator is 
$S[\varphi(t)]=S_{\rm ch}[\varphi(t)]+S_{\rm t}[\varphi(t)]$.
The first term represents the charging energy
\begin{equation}
	S_{\rm ch}[\varphi(t)] 
	= \int_{\rm K} \d t \left[ {C \over 2}\left( {\dot{\varphi}(t)
	\over e}\right)^2 + n_{\rm g} \dot{\varphi}(t)\right]\; .
\end{equation}
Electron tunneling is described by $S_{\rm t}[\varphi(t)]$, which, in the 
case of wide metallic junctions, is expressed by the simplest electron loop 
connecting two times,
\begin{equation}
	S_{\rm t}[\varphi(t)] = 2 \pi \i \sum\limits_{\rm r=L,R} 
	\int_{\rm K} \d t \int_{\rm K} 
	dt' \alpha^{\rm K}_{\rm r}(t,t') 
	\e^{i \varphi(t)} \e^{-i \varphi(t')} \; .
\end{equation}
The kernels $\alpha^{\rm K}_{\rm r}(t,t')=\alpha^\pm_{\rm r}(t-t')$ for
$t<t'$ ($t>t'$) depend on the order of the times along the Keldysh contour. 
Their Fourier transforms are \cite{SZ,schoell-schoen,koe-schoell-schoen1}
\begin{equation}
	\alpha^{\pm}_{\rm r}(\omega)=\pm\alpha_{\rm t,r} 
	\frac{\omega-\mu_{\rm r}}{\exp[\pm (\omega-\mu_{\rm r})/T] -1} \, .
\end{equation}
They are proportional to the dimensionless tunneling conductance
$\alpha_{\rm t,r}= h/(4 \pi^2 e^2 R_{\rm r})$ between the island and 
the leads r = L,R.

For large systems, the phase behaves almost like a classical variable while 
its conjugate variable, the charge, fluctuates strongly.
A natural basis is then the phase representation.
In the presence of strong Coulomb interaction, however, the situation is 
different: the phase underlies strong fluctuations while the time evolution
of the charge is almost governed by classical rates.
For this reason, it may be useful to change from the phase to the charge 
representation. The time evolution of the density matrix in a charge
representation depends on the propagator from $n_1$ forward to $n_1'$ 
and on the backward branch from $n_2'$ backward to $n_2$.
It is given by the matrix element of the reduced propagator 
\cite{koe-schoell-schoen1}
\begin{eqnarray}\label{Eq. propagator path}
	&&\!\!\!\!\!\!\!\!\!\!\!\!\!\!\!\!
	\Pi^{n_1,n_1'}_{n_2,n_2'} =\int d\varphi_1 \int d\varphi_1' 
	\int d\varphi_2' \int d\varphi_2 \,
	\e^{in_1\varphi_1} \e^{-in_1'\varphi_1'}
	\e^{in_2'\varphi_2'} \e^{-in_2\varphi_2}
	\\ \nonumber &&
	\int\limits^{\varphi_1,\varphi_1'}_{\varphi_2,\varphi_2'}
	{\cal D}[\varphi(t)] \int {\cal D} [n(t)] 
	\exp \left(-iS_{\rm ch}[n(t)] +iS_{\rm t}[\varphi(t)]+
	i \int_{\rm K} \d t \, n(t)\dot{\varphi}(t) \right) \, . \!\!\!\!\!\!\!
\end{eqnarray}
In the charge representation the charging energy is simply described by 
$S_{\rm ch}[n(t)]= \int_{\rm K} \d t \, E_{\rm C} 
\left[n(t)-n_{\rm g}\right]^2$.

\section{Expansion in the tunneling conductance}

A diagrammatic description is obtained by expanding the tunneling term
$\exp \left(iS_{\rm t}[\varphi(t)]\right)$ in the reduced propagator 
and integrating over $\varphi$.
Each of the exponentials $\exp[\pm i\varphi (t)]$ describes tunneling of an 
electron at time $t$.
These changes occur in pairs in each junction, r=L,R, and are connected by 
tunneling lines $\alpha^{\rm K}_{\rm r}(t,t')$.
Each term of the expansion can be visualized by a diagram.
Several examples are displayed in Fig.~\ref{fig1}.
The value of a diagram is calculated according the rules which follow from the
expansion of Eq.~(\ref{Eq. propagator path}) and are presented in detail in
Ref.~\onlinecite{koe-schoell-schoen1}.

The propagator from a diagonal state $n$ to another diagonal state
$n'$ is denoted by $\Pi^{n,n'}_{n,n'} = \Pi_{n,n'}$.
It is the sum of all diagrams with the given states at the ends and can be 
expressed by an irreducible self-energy part $\Sigma_{n,n'}$, defined as the 
sum of all diagrams in which any vertical line cutting through them crosses at 
least one tunneling line.
The propagator can be expressed as an iteration in the style of a Dyson 
equation, $\Pi_{n,n'} =	\Pi^{(0)}_n \delta_{n,n'} + \sum\limits_{n''} 
\Pi_{n,n''} \, \Sigma_{n'',n'} \, \Pi^{(0)}_{n'}$.
The term $\Pi^{(0)}$ describes a propagation in a diagonal state which does 
not contain a tunneling line.
The stationary probability for state $n$ follows from 
$P_n=\sum\limits_{n'}P^{(0)}_{n'}\,\Pi_{n',n}$ (in which $P^{(0)}_n$ is the 
initial distribution) and is {\it not} the equilibrium one if a bias voltage 
is applied.
Our diagram rules then yield
\begin{equation}\label{Eq. master}
	0 = \sum\limits_{n'} [ - P_n \Sigma_{n,n'} 
          + P_{n'} \Sigma_{n',n} ]\, .
\end{equation}
We recover the structure of a stationary master equation 
with transition rates given by $\Sigma_{n',n}$.
In general, the irreducible self-energy $\Sigma$ yields the rate of all 
possible correlated tunneling processes. 
We reproduce the well-known single-electron tunneling rates by evaluating all 
diagrams which contain no overlapping tunneling lines. 
Similarly cotunneling is described by the diagrams where two tunneling lines 
overlapping in time, as shown in Fig.~\ref{fig1}.

We calculate the current $I_{\rm r}$ flowing into reservoir ${\rm r = L,R}$ by 
adding a source term to the Hamiltonian and then taking the functional 
derivative of the reduced propagator with respect to the source.
The result 
$I_{\rm r}=-ie\int d\omega \left\{ \alpha_{\rm r}^+(\omega)C^>(\omega) + 
\alpha_{\rm r}^-(\omega)C^<(\omega) \right\}$ 
is expressed by the correlation functions 
$C^>(t,t')=-i{\langle \e^{-i\varphi(t)}\e^{i\varphi(t')}\rangle}$ and
$C^<(t,t')=i{\langle \e^{i\varphi(t')}\e^{-i\varphi(t)}\rangle}$ describing 
charge transfer at different times. These are related to the spectral density 
for charge excitations on the island by 
$2\pi \i A(\omega)=C^<(\omega)-C^>(\omega)$.

For sequential tunneling, the current reduces to
\begin{equation}
	I_{\rm r}={e\over h}4\pi^2\int d\omega\sum\limits_{\rm r'}
	{\alpha_{\rm r'}(\omega)\alpha_{\rm r}(\omega)\over\alpha(\omega)}
	A(\omega) [f(\omega - \mu_{\rm r'})-f(\omega-\mu_{\rm r})]
\label{resresult}
\end{equation}
with 
\begin{equation}
	A^{(0)}(\omega) = \sum\limits_{n=-\infty}^{\infty}
	[P_n+P_{n+1}]\delta(\omega-\Delta_n)
\end{equation}
and $\Delta_n=E_{\rm ch}(n+1)-E_{\rm ch}(n)=E_{\rm C}[1+2(n-n_{\rm g})]$.

At the minima of the Coulomb oscillations the system is in the Coulomb 
blockade regime, and cotunneling processes determine the conductance.
The second order processes are described by diagrams as shown in 
Fig.~\ref{fig1}. A careful analysis of our diagrammatic expansion not only
reproduces the known limits \cite{Av-Naz} but also provides the needed 
regularization of divergences.
We, furthermore, obtain new terms which are essential at the resonance.
These results will be presented in a forthcoming publication 
\cite{koe-schoell}.

At the resonance we have to include processes of arbitrary high order, 
since the process of resonant tunneling is essential.
For definiteness, we concentrate on situations where only two charge 
states, $n=0, 1$, need to be considered. 
This is the case when the energy difference of the two states 
$\Delta_0 \equiv E_{\rm ch}(1)-E_{\rm ch}(0)$, the bias voltage
$eV=eV_{\rm L}-eV_{\rm R}$, and the temperature $T$ are low compared to 
$E_{\rm C}$.
If, furthermore, we restrict ourselves to matrix elements of the density
matrix which are at most two-fold off-diagonal \cite{koe-schoell-schoen1}, 
we can evaluate -- in a conserving approximation -- the irreducible 
self-energy analytically.
The following results are derived in this limit.

Using the notations 
$\alpha_{\rm r}(\omega)=\alpha^+_{\rm r}(\omega)+\alpha^-_{\rm r}(\omega)$ and
$\alpha(\omega)=\sum\limits_{\rm r}\alpha_{\rm r}(\omega)$,
we find $P_0=\lambda_-$ and $P_1=\lambda_+$ with 
$\lambda_\pm=\int d\omega\, \alpha^\pm (\omega)|\pi(\omega)|^2$ and
\begin{equation}\label{sigma}
	\pi(\omega)=[\omega-\Delta_0-\sigma(\omega)]^{-1}
	\qquad\!\! , \qquad\!\!
	\sigma(\omega)= \int d\omega'\,{\alpha (\omega') \over 
	\omega-\omega'+i0^+}\, .
\end{equation}
Again, the current is given by Eq.~(\ref{resresult}), but the spectral 
density becomes
\begin{equation}\label{spectral density}
	A(\omega) = {\alpha(\omega) \over 
	[\omega-\Delta_0-\mbox{Re}\,\sigma(\omega)]^2
	+ [\mbox{Im}\,\sigma(\omega)]^2}\, .
\end{equation}
The following results depend on the parameter
\begin{equation}\label{alphat}
	\alpha_{\rm t} 
	=\sum\limits_{\rm r}\alpha_{\rm t,r} = \frac{h}{4\pi^2 R_{\rm t}},
\end{equation}
which also defines the parallel tunneling conductance 
$1/R_{\rm t}= \sum\limits_{\rm r} 1/R_{\rm r}$.
In lowest order in $\alpha_{\rm t} $
we have $A^{(0)}(\omega)=\delta(\omega-\Delta_0)$, 
and the classical result is recovered.
In general, quantum fluctuations yield energy renormalization and 
broadening 
effects, which enter in the spectral density via the complex self-energy 
$\sigma(\omega)$ given in Eq.~(\ref{sigma}).
In order to evaluate $\sigma(\omega)$ we introduce a Lorentzian cut-off
which we choose equal to $E_{\rm C}$ (since the energy difference to charge 
states which are not taken into account here is of the order of the charging 
energy).
In this case we find
\begin{equation}
	{\rm Re}\, \sigma(\omega)=- \sum\limits_{\rm r} \alpha_{\rm t,r}
	(\omega -\mu_{\rm r})\left[ 2\ln \left( {E_{\rm C}\over 2\pi T}
	\right) -2{\rm Re} \, \Psi \left(i \,{\omega- \mu_{\rm r}\over 2\pi T} 
	\right) \right]
\end{equation}
and ${\rm Im}\, \sigma(\omega)=-\pi \alpha(\omega)$.
The effect of the quantum fluctuations can be estimated from the spectral
density in the limits $T \gg eV,|\omega|$ or $eV \gg T,|\omega|$. 
Then, the spectral density is
\begin{equation}\label{renorm density}
	A(\omega) = {Z^2\alpha(\omega) \over 
	[\omega-Z\Delta_0]^2 + [\pi Z \alpha(\omega)]^2} \, ,
\end{equation}
with
\begin{equation}
	Z^{-1}=1+2\alpha_{\rm t} \ln({E_{\rm C} / \max \{eV/2,2\pi T\} }) .
\end{equation}
We observe a renormalization of $\Delta_0$ and $\alpha_{\rm t}$ by $Z$ and a 
broadening given by $\pi Z \alpha(\omega)$. From this result we 
conclude that lowest order perturbation theory is 
sufficient for 
$\alpha_{\rm t} \ln{(E_{\rm C} / \max\{eV/2,2\pi T\})}\ll 1$. 
At larger values, our results for resonant tunneling show clear deviations 
from sequential tunneling.

A pronounced signature of quantum fluctuations is contained in the 
differential conductance $G=\partial{I}/ \partial{V}$.
In Figs.~\ref{fig2} and \ref{fig3} we present our results for the 
differential conductance in the linear response regime ($V=0$).
They clearly display the effect of resonant tunneling:\\
-- For comparison, we show on the left hand side of Fig.~\ref{fig2} plots 
which are obtained from the master equation description of sequential 
tunneling, 
\begin{equation}
	{G(T,n_{\rm g}) \over G_{\rm as}} 
	= \frac{1}{\sum_n
	\exp \left[-{E_{\rm C}\over T}(n-n_{\rm g})^2\right]}
	\sum_{n=-\infty}^\infty 
	\exp \left[-{E_{\rm C}\over T}(n-n_{\rm g})^2 \right] 
	\frac{{E_{\rm C} \over T} (1+2(n-n_{\rm g}))}
	{\exp\left[{E_{\rm C}\over T}(1+2(n-n_{\rm g}))\right] -1} \, .
\end{equation}
The asymptotic high-temperature conductance is 
$G_{\rm as}=1/(R_{\rm L}+R_{\rm R})$. 
At low temperatures, when processes involving only two charge states dominate,
the maximal classical conductance saturates at one half of the asymptotic 
conductances at high temperatures.
The width of the peaks scale linearly with temperature.\\
-- The situation changes when resonant tunneling processes are taken into 
account (see the plots on the right hand side of Fig.~\ref{fig2}). 
The maximal conductance and the peak width are renormalized by $Z$ and 
$Z^{-1}$ which depend logarithmically on temperature.
For this reason, the conductance peak does not reach one half of the high 
temperature limit and decreases with lower temperatures, while the peak width 
is increased compared to the lowest order perturbation theory result.
For an estimate of the maximal conductance, we use can the spectral density 
in the form of Eq.~(\ref{renorm density}) and perform the integral 
Eq.~(\ref{resresult}) analytically,
\begin{equation}
	{G_{\rm max}(T)\over G_{\rm as}} 
		\approx Z\left[{1\over 2}-{1\over \pi}
	\arctan \left( { (\pi Z \alpha_{\rm t})^2 -1 \over
	2\pi Z \alpha_{\rm t}} \right) \right] \, .
\end{equation}
(The results shown in Fig.~(\ref{fig3}), however, were obtained by numerical
analysis based on Eq.~(\ref{spectral density}).)

Recent experiments \cite{est,Kuz} in systems with junctions with small 
barriers show, indeed, a broadening and decreasing height of the linear 
conductance peaks, which cannot be explained by thermal smearing and
qualitatively agrees with our theory.

The effects of quantum fluctuations are even more pronounced 
in the nonlinear differential conductance when the transport voltage 
dominates over temperature. In Fig.~\ref{fig4} we compare the 
results of perturbation theory and resonant 
tunneling at $T=0$ assuming that for $eV<2E_{\rm C}$ only two charge states 
$n=0,1$ are involved.\\
-- The sequential tunneling result for a symmetric transistor 
with $\alpha_{\rm t,L}=\alpha_{\rm t,R}$ and $C_{\rm L}=C_{\rm R}$ is
\begin{equation}
	{G(V,n_{\rm g}) \over G_{\rm as}} 
	= 2{E_{\rm C}^2(1-2n_{\rm g})^2+(eV)^2/4 \over (eV)^2} 
	\,\, \Theta \left( {eV\over 4 E_{\rm C}}- \left| n_{\rm g}-{1\over 2}
	\right| \right) \; .
\label{mx}
\end{equation}
As a function of $n_{\rm g}$ it shows a series of structures of width $CV/e$ 
with vertical steps at its edges. The width scales linearly with bias 
voltage.\\
-- Resonant tunneling leads to a renormalization of the height and width by
$Z$ and $Z^{-1}$ respectively, which depends now logarithmically on the 
voltage (see Fig.~\ref{fig4}).
For this reason, the height of the structure is below the sequential tunneling
result and further decreases at lower voltages, while the width is enhanced.
Furthermore, the sharp edges are smeared out even in the absence of thermal 
fluctuations (since $T=0$).

\section{strong tunneling}

If the junction conductance is high and hence the fluctuations in the
charge are strong the phase representation outlined above is a more
suitable starting point for the analysis of the problem.
It turns out that the dimensionless conductance appears in the form
\begin{equation}
	\tilde \alpha _{\rm t} =\frac{h}{4e^2R_t} = \pi^2 \alpha _{\rm t} \;,
\label{alphatilde}
\end{equation}
which differs from the expansion parameter $\alpha _{\rm t}$ of the
weak tunneling expansion by a factor $\pi^2$.
The real-time path-integral technique
discussed above provides an expression for the reduced density matrix $\rho
(\varphi_1,\varphi_2)$. If the island charge can vary continuously the
density matrix $\rho _{\rm c}(\varphi_1,\varphi_2)$ is given by 
Eq.~(\ref{continuous}). It obeys the standard normalization condition 
$\int\limits_{-\infty }^{+\infty }d\varphi \rho_{\rm c}(\varphi ,\varphi
)=1$ and its time evolution is governed by the action
 $S[\varphi (t)] = \int_K \d t {\frac{C}{2}}
\left( {\frac{\dot {\varphi}(t)}{e}} \right)^2 
+ S_{\rm t}[\varphi (t)]$. In this case the charging energy does not
depend on the gate charge. In the SET transistor, 
another physical situation is realized, where the island charge is discrete
and quantized in units of $e$. This situation is described by the density
matrix $\rho_{\rm d}(\varphi_1,\varphi_2)$, Eq.~(\ref{discrete}), 
with a compact phase variable~\cite{SZ}. 
The normalization of the density matrix (\ref{discrete}) is
given by 
$\int\limits_{-\pi }^{\pi }d\varphi \rho _{\rm d}(\varphi ,\varphi )=1$.
It is sensitive to the gate charge. The comparison of Eq.~(\ref{continuous}) 
and (\ref{discrete}) shows the following relation 
\begin{equation}
	\rho_{\rm d}(\varphi,\varphi)
		=A^{-1}\sum\limits_{m_1,m_2} \e^{i2\pi n_{\rm g}(m_1-m_2)}
		\rho_{\rm c} (\varphi+2\pi m_1,\varphi+2\pi m_2) \;.  
\label{m}
\end{equation}
Here $A$ is a normalization factor 
$A= \sum\limits_m\int\limits_{-\infty }^{+\infty }d\varphi \e^{ i2\pi n_{\rm g}
m} \rho _{\rm c}(\varphi +2\pi m,\varphi )$.
The relation (\ref{m})  can also be used to 
establish the connection between the expectation
values of physical quantities  for systems with discrete and
continuous charge distributions.
The expectation value of an operator $\hat O(\hat \varphi)$, 
of the discrete-charge system, which is $2\pi $-periodic in $\varphi$ is
\begin{eqnarray}
	\left\langle \hat O\right\rangle _{\rm d} 
	&=&	\int\limits_{-\pi }^{\pi }d\varphi
	O(\varphi) \rho_{\rm d}(\varphi ,\varphi )  
	=\sum\limits_m\int\limits_{-\infty }^{+\infty}
	d\varphi O(\varphi ) \e^{i2\pi n_{\rm g} m} 
	\rho_{\rm c} (\varphi -2\pi m,\varphi ) \nonumber \\
	&=&\frac{\sum\limits_m\left\langle \hat O(\hat \varphi) 
	\e^{i2\pi (n_{\rm g}-\hat n)m} \right\rangle _{\rm c}} 
	{\sum\limits_m\left\langle \e^{i2\pi (n_{\rm g}-\hat n) m}
	\right\rangle _{\rm c}}.  \label{av1}
\end{eqnarray}
Here we used an obvious identity 
$\int d\varphi (...) \rho_{\rm c}(\varphi -2\pi m,\varphi ) 
	= \langle (... ) \e^{-2\pi \i \hat n m}\rangle _{\rm c} $. 

Now we are ready to evaluate the tunneling current through a SET
transistor. We first derive an expression for the expectation value of
the current and then evaluate it with the aid of 
Eq.~(\ref{av1}). The first part of this program will be carried out
within the quasiclassical Langevin equation approach 
\cite{Eckern,S,GZ92,GZ96,Od} derived
under the assumption that fluctuations of the phase variable are
 weak. This assumption is justified if the fluctuations of the
charges are strong. 

In a semiclassical approximation we find for the
current through the the left and the right junctions ${\rm r=L,R}$
(see Refs.~\onlinecite{Eckern,GZ92} for further details)
\begin{equation}
	i_{\rm r}=C_{\rm r}\frac{\ddot \varphi _{\rm r}}e 
	+\frac 1{R_{\rm r}}\frac{\dot \varphi_{\rm r}}{e}
	-\tilde \xi_{r}(\varphi ) \;.
\label{eq}
\end{equation}		
It depends on the {\it fluctuating} voltage differences
 across the junctions, $\dot \varphi_{\rm r}/e = U - V_{\rm r}$ 
for ${\rm r} = 1,2$. Here $U$ is the voltage of the island.
Each current is the sum of displacement current on the capacitor, a 
deterministic tunneling current and a shot noise contribution.
The latter can be expressed as a state-dependent noise 
in the form~{\cite{Eckern}
\begin{equation}		
		\tilde \xi_{\rm r}
	=\xi_{\rm r1}(t) \cos \varphi_{\rm r} 
	+\xi_{\rm r2}(t)\sin \varphi_{\rm r} \; ,
\label{eqn}
\end{equation}
where $\xi_{{\rm r}i}$ ($i=1,2$) are two independent
Gaussian stochastic variables with correlators  
\begin{equation}
	\left\langle \xi _{{\rm r}i}(0)\xi _{{\rm r}^{\prime}j}(t)
	\right\rangle 
	= \delta_{\rm r,r^{\prime}} \delta_{i,j}\frac 1{R_{\rm r}} 
	\int \frac{d\omega }{2\pi } \e^{ i\omega t} \omega 
	\coth \left( \frac \omega {2T}\right) .
\end{equation}

We consider situations where the external impedance is negligible.
In this case the phases $\varphi _{\rm r}$ are linked
to the transport voltage $V$ by $\dot \varphi
_{\rm L}-\dot \varphi_{\rm R}=eV$. We can further assume a symmetric
bias $V_{\rm L}=-V_{\rm R}=V/2$. 
In this case the voltage on the island can be expressed as
$U = (\dot \varphi _{\rm L}+ \dot \varphi
_{\rm R})/2e=V_{\rm g}+ \dot \varphi_{\rm g}/e$.
Here $\dot{\varphi}_{\rm g}/e$ is the voltage across the gate capacitance.
Finally, charge conservation implies
$i_{\rm L}+ i_{\rm R}=-C_{\rm g} \ddot \varphi _{\rm g}/e$.
Combining these relations with (\ref{eq})  we arrive, after averaging 
over the stochastic variables $\xi $, at the expression for the 
expectation value of the current 
\begin{equation}
	I=\langle i_{\rm L}\rangle 
	=\frac{V - 
	 R_{\rm L}\left\langle \tilde \xi_{\rm L}\right\rangle_{\rm d}
	-R_{\rm R}\left\langle \tilde \xi_{\rm R}\right\rangle_{\rm d}}
	{R_{\rm L}+R_{\rm R}} \; .
\label{vax1}
\end{equation}
This expression will be evaluated further with the aid of
 relation (\ref{av1}). 
If the fluctuations of the charge can be treated as Gaussian the
contribution of the $m$-th winding number to the expectation
value (\ref{av1}) can be estimated as 
\begin{equation}
	\left\langle \dots \e^{i2\pi \hat nm}
	\right\rangle_{\rm c}
	\sim \exp \left(
	-2\pi ^2\left\langle \delta \hat n^2\right\rangle_{\rm c} 
	m^2\right) .
\label{estimation}
\end{equation}
Thus provided that the charge fluctuations are not small 
$\left\langle \delta \hat n^2\right\rangle_{\rm c} \gtrsim 1$ 
it is sufficient to retain in the
expression (\ref{av1}) only terms with winding numbers
$m=0,\pm 1$. In this approximation we obtain \cite{GZ96}
\begin{equation}
	\left\langle \tilde \xi _{\rm r}\right\rangle _{\rm d}
	=\frac{\ \left\langle \tilde \xi_{\rm r}\right\rangle_{\rm c} 
	+2\left\langle \tilde \xi _{\rm r} 
	\cos (2\pi (\hat n -n_{\rm g}))\right\rangle_{\rm c} }
	{1+2\left\langle 
	\cos (2\pi (\hat n-n_{\rm g}))\right\rangle_{\rm c}} \;.
\label{average}
\end{equation}

In the quasiclassical limit considered here the further analysis
requires standard noise averaging of the solutions of Eqs. (\ref{eq}). 
As these equations are nonlinear in the phase, the exact solution
cannot be constructed in
general. In a wide parameter range, however, it is sufficient
to proceed perturbatively in the noise terms.
Making use of (\ref{vax1}), (\ref{average}) and assuming the phase
fluctuations to be small $|\delta \varphi |\lesssim \pi $ we arrive at the
following expression for the current 
\begin{eqnarray}
	I(V)=&&G_{\rm as}V
	-I_0(V)
	\nonumber \\
	&&-\e^{-F(T,V)}\left[ \left[ I_1(V)-2I_0(V)\right]
	\cos \left( \frac{2\pi Q_{\rm av}(V)}e\right) 
	+I_2(V)\sin \left( \frac{2\pi
	Q_{\rm av}(V)}e\right) \right] ,  \label{vax2}
\end{eqnarray}
Here
\begin{equation}
	Q_{\rm av}=\langle e(\hat n-n_{\rm g})\rangle_{\rm c} 
	=\frac{C_{\rm L}R_{\rm L}-C_{\rm R}R_{\rm R}}{R_{\rm L}
	+R_{\rm R}}V+C_{\rm g}V_{\rm g}
\label{Qav}
\end{equation} 
is the average charge of the metallic island. We further introduced the
integrals
\begin{equation}
	I_0(V)=\frac{2e}{\pi (R_{\rm L}+R_{\rm R})}
	\int\limits_0^\infty \d t\left( \frac{\pi T}{\sinh \pi
	Tt}\right)^2
	\e^{-W(t,V)}K(t)\cos \left( \frac{e(R_{\rm R}-R_{\rm L})Vt}{%
2(R_{\rm R}+R_{\rm L})}\right) \sin \left( \frac{eVt}2\right) ,  \label{I0}
\end{equation}
\begin{eqnarray}
\begin{array}{c}
I_1(V) \\ 
I_2(V)
\end{array}
&=&\frac{4e}{\pi (R_{\rm L}+R_{\rm R})}\int\limits_0^\infty \d t\left(
\frac{\pi T}{\sinh \pi Tt}\right) ^2\e^{-W(t,V)}\left[ K(t)
\begin{array}{c}
\cosh  \\ 
\sinh 
\end{array}
(u(t,V))\right.   \nonumber \\
&&\left. +4\pi C\dot K(t)
\begin{array}{c}
\sinh  \\ 
\cosh 
\end{array}
(u(t,V))\right] 
\begin{array}{c}
\cos  \\ 
\sin 
\end{array}
\left( \frac{e(R_{\rm R}-R_{\rm L})Vt}{2(R_{\rm R}+R_{\rm L})}\right) \sin 
\left( \frac{eVt}2\right) \; ,  \label{I12}
\end{eqnarray}
and we defined $K(t)=R_t\theta (t)(1-\exp (-t/R_tC))$,  
$1/R_t=1/R_{\rm L}+1/R_{\rm R}$
and
\begin{eqnarray}
	W(t,V) = -\frac{e^2}{2\pi }\int\limits_{-\infty }^{+\infty}
	dt_1\int\limits_{-\infty }^{+\infty }dt_2
	\left( \frac{\pi T}{\sinh \pi T(t_1-t_2)}\right)^2
	K(t_1,t)K(t_2,t) 
	\nonumber \\
	\sum\limits_{\rm r=L,R}
	\frac 1{R_{\rm r}}\cos \left( \frac{eR_{\rm r}V(t_1-t_2)}{%
R_{\rm L}+R_{\rm R}}\right) , \label{W} 
\end{eqnarray}
with $K(t',t)\equiv K(t')-K(t'-t)$. A principal value of the time integrals
in (\ref{I12}), (\ref{W}) should be taken where needed.

The function $F(T,V)=2\pi^2\langle\delta n^2\rangle $  determines 
the temperature and voltage dependence of the charge fluctuations
in the Gaussian approximation.
It is given by an expression similar to $W(t,V)$ (\ref{W}) 
with the substitution $K(t_1,t)K(t_2,t) \to
(4\pi^2C^2/e^4)\dot K(t_1)\dot K(t_2)$. Analogously $u(t,V)$ is defined
by (\ref{W}) after the substitution $K(t_1,t) \to
-(4\pi C^2/e^2)\dot K(t_1)$.
	
We can simplify these expressions observing that in the
limit of sufficiently high temperatures and/or voltages 
\begin{equation}
 	\max \{eV,T\} \gg w_0=\frac{2\tilde \alpha _{\rm t}
	E_{\rm C}}{\pi^2} \exp (-2\tilde \alpha _{\rm t}+\gamma ),\;\;\;%
 \label{req}
\end{equation}
(here $\gamma =0.5772...$ is Euler's constant) 
the results can be simplified further, since the
time integration in (\ref{W}) is effectively cut off at short times.
Since $W(t=\min \{1/T,1/eV\},V)\ll 1$ we can set in the parameter 
range (\ref{req}) in leading order approximation $W(t,V)=0$. With this
simplification the above integrals can be evaluated
analytically. We obtain 
\begin{eqnarray}
I_0(V) = \frac{eR_0}{R_{\rm L}+R_{\rm R}}
	\sum_{\rm r =L,R}\left\{ \frac{eR_{\rm r}V}{\pi
(R_{\rm L}+R_{\rm R})}\left[ {\rm Re}\Psi \left( 1+\frac 1{2\pi
TR_0C}- \i \frac{eR_{\rm r}V}{2\pi T(R_{\rm L}+R_{\rm R})}\right) 
\right. \right. \nonumber \\ \left. \left.
-{\rm Re}\Psi \left( 1-
\i \frac{eR_{\rm r}V}{2\pi 
T(R_{\rm L}+R_{\rm R})}
\right) \right] \right\}   \nonumber \\
-\frac 1{\pi R_0C}\sum\limits_{\rm r =L,R}{\rm Im}\Psi \left(
1+\frac 1{2\pi TR_0C}- \i \frac{eR_{\rm r}V}{2\pi T(R_{\rm L}
+R_{\rm R})}\right) \;,  \label{I0an}
\end{eqnarray}
and
\begin{eqnarray}
F(T,V) &=&F(0,0)+\frac{2\pi ^2CT}{e^2}+\frac{2\pi }{e^2R_0}
	\ln \left( \frac 1 {2\pi TR_0C}\right) -  \nonumber \\
	& &\frac{2\pi }{e^2}\sum\limits_{\rm r =L,R}{\rm Re}
	\left[ \frac 1{R_{\rm L}}
	\left( 1-\i \frac{eR_{\rm r}R_0C}{R_{\rm L}
	+R_{\rm R}}V\right) \Psi \left( 1+\frac1{2\pi TR_0C} 
	-\i \frac{eR_{\rm r}V}{2\pi T(R_{\rm L}
	+R_{\rm R})}\right) \right] \; .  \label{Fan}
\end{eqnarray}
Here $\Psi (x)=\Gamma ^{\prime }(x)/\Gamma (x)$ is the digamma
function.  

The last expression determines 
the temperature and voltage dependence of the charge fluctuations 
$\langle \delta n^2\rangle$ at not too low $T$ and/or $V$. 
At $T=0$ and $V=0$ the integral over time in  $F(T,V)$ 
diverges logarithmically at high frequencies. 
This divergence indicates a failure of the quasiclassical 
Langevin equation in this limit. 
The problem can be cured by observing that in thermodynamic
equilibrium (zero voltage) in Gaussian approximation 
\begin{equation}
	\e^{-F(0,0)}
	=\left\langle \cos(2\pi \hat n)\right\rangle = \frac{\int d\varphi 
	\rho _{\text{eq} }(2\pi +\varphi ,\varphi )}{\int
	d\varphi \rho _{\text{eq}}(\varphi ,\varphi )}  \;.
\end{equation}
The expectation value involving the equilibrium density matrix 
 can be evaluated \cite{GZ96} with the result
\begin{equation}
	F(0,0)\simeq 2\tilde \alpha _{\rm t}\;.
\label{cutoff}
\end{equation}

The functions $I_1(V)$ and $I_2(V)$ cannot be evaluated analytically even in
the limit (\ref{req}). Due to a fast decay of the exponential factor $\exp
[-F(T,V)]$ in (\ref{vax2}) with increasing $V$ and $T$ it is sufficient to
evaluate $I_1$ and $I_2$ in the low voltage and temperature limit. In this
limit the integral (\ref{I12}) reduces to
\begin{equation}
I_1(V)-2I_0(V)\approx gG_{\rm as}V ,\;\;\;
g=\frac{1.22}{\tilde \alpha
_{\rm t}}+11.29.  \label{I1an}
\end{equation}
whereas the function $I_2(V)$ turns out to be small $I_2(V)\sim V^2\approx 0$
and will be neglected below.

We thus arrive at the following result for the I-V characteristics of a SET
transistor 
\begin{equation}
I(V)=G_{\rm as} V -I_0(V)
-gG_{\rm as}V\e^{-F(T,V)}\cos \left( \frac{%
2\pi Q_{\rm av}(V)}e\right) \;.  
\label{vax3}
\end{equation}
The current is reduced below the classical result $G_{\rm as} V$ by an
amount $I_0(V)$ and is modulated in a  periodic way by the gate
voltage. In the limit considered the modulation is a pure
cos-modulation. The result (\ref{vax3})
also describes the oscillatory behavior of the current as a function
of the transport voltage, which is usually referred to as a ``Coulomb
staircase''. The amplitude of these oscillations decays exponentially with
increasing voltage and temperature. We also recover the fact that the Coulomb
staircase is pronounced only in asymmetric SET transistors. In a
symmetric case the transport voltage drops out from the expression for the
gate charge (\ref{Qg}). The I-V characteristics (\ref{vax3}) is 
depicted in Fig.~\ref{fig5}
for different temperatures and values of the gate charge.

The linear conductance of a SET transistor can be easily derived from Eq.~(\ref
{vax3}) in the limit  $V\rightarrow 0$. We find 
\begin{equation}
	{\frac{G(T)}{G_{\rm as}}}
	= 1-f(T)-g\e^{-F(T,0)} \cos \left(2\pi n_{\rm g}\right) \;,  
\label{G}
\end{equation}
where $n_{\rm g}=C_{\rm g}V_{\rm g}/e$ and 
\begin{equation}
	f(T)=\frac 1{2\tilde \alpha _{\rm t}}
	\left[ \gamma +\frac{2\tilde \alpha _{\rm t}E_{\rm C}}
	{\pi ^2T}\Psi ^{^{\prime }}
	\left( 1+\frac{2\tilde\alpha _{\rm t}E_{\rm C}}{\pi
	^2T}\right) 
	+\Psi \left( 1+ \frac{2\tilde\alpha_{\rm t}
	E_{\rm C}}{\pi ^2T}\right) \right] \;. 
\label{f(T)}
\end{equation}
These results are
displayed in Figs.~\ref{fig6} and ~\ref{fig7}
in the temperature range $T \gtrsim 10w_0$, where we estimate  the 
approximations used above to be justified.  

In the high-temperature limit the
conductance becomes independent of the gate charge, but due to charging
effects it is still reduced below  the asymptotic value by
\begin{equation}
	{\frac{G(T)}{G_{\rm as}}}
	=1-\frac{E_{\rm C}}{3T}+ \left(\frac{6\zeta (3)}{\pi ^4}
\tilde \alpha _{\rm t} +\kappa \right)\left(\frac{E_{\rm C}}T\right)^2-...
\label{asympt}
\end{equation}
For high temperatures this expression is valid for all (including small)
values of $\tilde \alpha_{\rm t}$. The first nontrivial 
term in this expansion does not depend on $%
\tilde \alpha_{\rm t}$. The coefficient of the square term also contains an 
$\tilde \alpha_{\rm t}$-independent contribution $\kappa$. 
Within the approximation $W=0$ used here we have 
$\kappa =0$. An improved approximation
is obtained by expanding in $W(t,V)$, or alternatively by
treating the general expression for the
system conductance perturbatively in $\tilde \alpha_{\rm t}$ and then
expanding  in $E_{\rm C}/T$. This procedure yields $\kappa =1/15$,
which for large $\tilde \alpha_{\rm t}$ can be neglected 
against the first
term $6\zeta (3)\tilde \alpha_{\rm t} /\pi^4$.

At lower temperatures the conductance is further
suppressed by charging effects and it can be modulated by the gate 
charge $Q_{\rm g}$. In the figures
the minimum and maximum conductance values are presented 
corresponding to $Q_{\rm g}=0$ and $Q_{\rm g}=e/2$, as well as 
the $Q_{\rm g}$-averaged conductance.
The modulation with 
$Q_{\rm g}$ becomes more pronounced as the temperature is lowered,
however, it is exponentially 
suppressed with increasing $\tilde \alpha_{\rm t}$ (cf. Figs.~\ref{fig6}
and ~\ref{fig7}). For $\tilde \alpha_{\rm t} \gtrsim 4$ the modulation
effect can hardly be resolved while the overall suppression of the system
conductance $G$ is very pronounced.

Although the validity of the Langevin description
description is restricted to high temperatures and/or voltages,
$T,eV \gg w_0$, the validity range rapidly expands 
as $\tilde \alpha_{\rm t}$ increases. 
E.g. for the parameters
 $E_{\rm C} \sim 1$~K and $\tilde \alpha_{\rm t} \approx 2$, 
we get $w_0$ in the range 15 mK. 
Further increase of $\tilde \alpha_{\rm t}$ rapidly brings $w_0$ below 
1 mK. Therefore we can conclude that in the strong tunneling regime 
$\tilde \alpha_{\rm t} >2\div 3$ our theory covers
the experimentally accessible temperatures. Indeed a quantitative 
agreement without fitting parameters exists 
between our results (\ref{G}-\ref{asympt}) and those of the  
Saclay group \cite{est} in the high temperature regime.
For lower temperatures
the quasiclassical Langevin equation approach
can be applied only to sample 4 of Ref.~\onlinecite{est}
with $\tilde \alpha_{\rm t} \simeq 1.8$.
Other samples studied in  Ref.~\onlinecite{est}
 have substantially lower conductance, 
and their low-temperature
behavior should be described by the expansion in $\alpha_{\rm t}$
presented in Section 3.

\section{Discussion}

In a number of earlier papers \cite{Guinea,PZ91,ZP93,Falci} 
the combination of
charging and strong tunneling effects in metallic junctions has been
analyzed within imaginary time approaches. In the limit of strong
tunneling, $\tilde \alpha_{\rm t} \gg 1$,
a renormalization group equation for $\tilde \alpha_{\rm t}$ can be derived
\cite{Guinea,SZ}
\begin{equation}
	d\tilde \alpha_{\rm t}/d\ln \omega_{\rm c}
	=\beta (\tilde \alpha_{\rm t} ) \;, \label{RG}
\end{equation}
where in the lowest order in $\tilde \alpha_{\rm t}$ one has  
$\beta (\tilde \alpha_{\rm t} )=1/2$.
Already this scaling approach captures the tendency of the effective
junction conductance to decrease with decreasing $T$ due to charging
effects. In order to see that one should proceed with 
scaling from $\omega_{\rm c}
\sim E_{\rm C}$ to $\omega_{\rm C} \sim T$ and 
identify the (dimensionless) junction
conductance with the renormalized value 
$\tilde \alpha_{\rm t} (\omega_{\rm c} \sim T)$.
This approach is sufficient for strong tunneling at high temperatures, 
namely if the final renormalized tunneling conductance still
satisfies $\tilde \alpha_{\rm t} (\omega_{\rm c} \sim T) \gtrsim 1$. 
In general the strong tunneling approach  may lead to a small
renormalized conductance such that (\ref{RG}) ceases to be
valid. For weak tunneling other scaling approaches, derived in an
expansion in the tunneling conductance and equivalent to what
we described in Section III, can be applied. In this situation, Falci {\sl
et al.}~\cite{Falci} suggested a 2-stage scaling procedure,
where the renormalized conductance after the strong tunneling
rescaling was used as an entry parameter for the weak tunneling
scaling.

Various theoretical approaches led to the conclusion the strong
electron tunneling  $\tilde \alpha_{\rm t} \gg 1$
reduces the charging energy, i.e. the effective
capacitance is renormalized. Panyukov and Zaikin \cite{PZ91}
treated the problem by means of instanton techniques. They concluded
that electron tunneling  affects both the scale and the 
functional dependence of the ground state energy $E(Q_{\rm g})$. 
At not too low temperatures $T \gtrsim w$ they find
\begin{equation}
E(Q_{\rm g})=-\frac{w}2\cos (\frac{2\pi Q_{\rm g}}e)  \label{EQ}
\end{equation}
with~\cite{PZ91} 
\begin{equation}
	w=\frac{32\tilde\alpha_{\rm t}E_{\rm C}}{\pi^2}
	\exp (-2\tilde \alpha_{\rm t} +\gamma). 
\label{w}
\end{equation}
A similar result, differing only in the numerical coefficient,
has been obtained in a semiclassical analysis of the effective 
action~\cite{Falci}.

At lower temperatures the form of the lowest energy band 
$E(Q_{\rm g})$ turns out
to be even more complicated \cite{PZ91,Falci} 
and the $\tilde\alpha_{\rm t}$-dependence
of the prefactor of the expression for $w$ changes from linear in $\tilde 
\alpha_{\rm t}$ for $T >w$ to 
quadratic in $\tilde \alpha_{\rm t}$ for $T =0$.
Instanton techniques~\cite{PZ91} yield
\begin{equation}
	E_{\rm C,eff} \propto 
	\alpha_{\rm t}^2E_C\exp (-2\tilde \alpha_{\rm t})\;.
\label{Ceff}
\end{equation}
The exponential dependence on
$\alpha_{\rm t}$ has been  confirmed by 
renormalization group arguments~\cite{Guinea,SZ,Falci} as well as
Monte Carlo methods~\cite{Falci,WEG}. 
The prefactor remains a point of controversial discussions 
in the literature ~\cite{WEG}. Irrespective of this detail
an important consequence of the strong capacitance renormalization
for $\alpha_{\rm t} \gtrsim 1$ is the exponential reduction  of the 
temperature range where charging effects are observable.

With the aid of relations (\ref{EQ}), (\ref{w}) we can derive 
the first order  correction in $1/\tilde\alpha_{\rm t}$
in the renormalization group equation (\ref{RG})  \cite{Fn} 
\begin{equation}
\beta (\tilde \alpha_{\rm t} )=1/2+1/4\tilde\alpha_{\rm t}.  \label{first}
\end{equation}
This result has also been derived by direct RG methods~\cite{zwerger}.

A consequence of the renormalization group approach (\ref{RG})
has been pointed out in Ref. \onlinecite{PZ91}. It relies on the  
{\it assumption} that the system linear conductance 
is determined by the renormalized value 
$\tilde \alpha_{\rm t}(\omega_{\rm c} \sim T)$ as 
\begin{equation}
	G=\frac{2e^2}{\pi \hbar} 
	\tilde\alpha_{\rm t} (\omega_{\rm c} \sim T) \;.
\label{grenorm}
\end{equation}
Combining the above scaling approach, the high temperature
expansion (\ref{asympt}) (with $\kappa =0$), and the expression
(\ref{first}) for $\beta$  to first order in $1/\alpha_{\rm t}$ 
we get for the 
$Q_{\rm g}$-averaged conductance 
\begin{equation}
\frac G{G_{\rm as}}=1-\frac 1{2\tilde\alpha _{\rm t}}\left\{ \ln 
	\left( 1+\frac{4\tilde\alpha _{\rm t}^2}
	{ 3(1+2\tilde\alpha _{\rm t})}\frac{E_{\rm c}}T\right) +\ln \left[ 
	1+\frac 1{%
	2\tilde\alpha _{\rm t}}\ln \left( 1+\frac{4\tilde\alpha _{\rm t}^2}
	{3(1+2\tilde\alpha _{\rm t})%
	}\frac{E_{\rm C}}T\right) \right] \right\} . \label{RGcond}
\end{equation}
Although the above scaling approach to the conductance calculation is
intuitively attractive (and the result (\ref{RGcond}) fits reasonably 
with the available experimental data \cite{est,Kuz})
it has to be stressed that it depends on the unproven
 assumption (\ref{grenorm}).

In contrast, the real-time path-integral techniques presented here are free
from this ambiguity and allow for a direct evaluation of the 
$I$-$V$ characteristics and the system conductance. 
We note, furthermore, that  the results obtained
within the real and imaginary time methods are consistent with each other. 
E.g. the renormalization of the effective energy difference between
the two lowest charge states, derived in Ref.~\onlinecite{Falci}, is
contained in the self-consistent solution presented in Section 3. 
Furthermore, comparing the expressions
for $w_0$ (\ref{req}) and the bandwidth $w$ (\ref{w}) we immediately see 
that these two parameters coincide up to a numerical coefficient:
$w= 16w_0$. This means the requirement for the validity
of the quasiclassical Langevin equation  (\ref{req}) roughly
coincides with the requirement that the temperature (or 
voltage) is  larger than the effective bandwidth $w$.

Still no quantitative theory for the conductance at lower temperatures and
not too low values  $\tilde\alpha_{\rm t} \gtrsim 1$ 
has been provided. Although the two  limiting descriptions
 presented here do not allow for a quantitative
description of this parameter range it satisfactory to notice
that both  show the same qualitative trend in this range.

Another question of interest is the conductance
at very large $\tilde \alpha_{\rm t} \gg 1$ and very low $T \lesssim w_0$.
In the limit $\tilde \alpha_{\rm t} \gg 1$ the conductance oscillations with 
$Q_{\rm g}$ are exponentially small (cf. (\ref{G})). Then for all $Q_{\rm g}$
from (\ref{G},\ref{cutoff}) we have 
\begin{equation}
G(T\approx w_0)/G_{as} \simeq b/\tilde\alpha_{\rm t}, \;\;\; b \sim 1.
\label{b}
\end{equation} 
Thus we can {\it conjecture} that the 
low temperature maximum conductance of a SET transistor is 
{\it universal} in the limit of large $\tilde \alpha_{\rm t}$ 
being of the order of the inverse quantum resistance
unit $2e^2 /\pi \hbar $. This conjecture is also 
consistent with the scaling analysis of 
Refs.~\onlinecite{Guinea,PZ91,Falci} combined with the results of Section 3. 
Starting from large $\tilde \alpha_{\rm t} \gg 1$ we first use the
renormalization group procedure (\ref{RG},\ref{first}) 
which should be cut at
$\tilde \alpha_{\rm t} (\omega_{\rm c}) \sim 1$.
In the second stage we  expand in $\alpha_{\rm t} \approx 1/\pi^2$ 
as described in Section 3  -- starting
with the renormalized value instead of the bare one. 
Apart from logarithmic corrections we thus
arrive at the maximum conductance of order of the inverse quantum resistance,
no matter how large the initial conductance is.

\section{Conclusions}

In this paper we have described single-electron tunneling in systems with
strong charging effects beyond perturbation theory in the tunneling
conductance. For this purpose we considered the real-time evolution of the
reduced density matrix of the system. We presented two approximation
schemes: \\In the first part, valid for not too strong tunneling, $\alpha_{%
{\rm t}} \lesssim 1$, we presented a systematic diagrammatic expansion,
which allowed us to identify the different contributions, sequential
tunneling, inelastic cotunneling and inelastic resonant tunneling. When we
restricted ourselves to diagrams corresponding to maximally two-fold
off-diagonal matrix elements of the density matrix we can formulate a
self-consistent resummation of diagrams. At low temperatures we,
furthermore, can restrict our attention to two consecutive charge states. In
this limit, there exist no crossing diagrams, and we can evaluate the
summation in closed form. The most important results are a renormalization
of system parameters and a life-time broadening of the conductance peaks.
These two approximations are justified for tunneling conductances satisfying 
$\alpha_{{\rm t}} \ln{(E_{{\rm C}} / \max\{eV/2,2\pi T\})}\lesssim 1$ and
allow for a qualitative analysis of the system conductance also for larger
values of $\alpha_{{\rm t}}$.

In the second part of the paper we developed an alternative approach based
on quasiclassical Langevin equations for the junction phase $\varphi$. This
approach assumes that fluctuations of the phase are small and that the noise
can be treated perturbatively. This is a suitable approximation for large
values $\tilde \alpha_{{\rm t}} =\pi^2 \alpha_{{\rm t}}$ or in the high 
temperature limit. For weak
tunneling $\tilde \alpha_{{\rm t}} \lesssim 1$ this scheme turns out to be
justified only for high temperatures and/or voltages max$(T,eV) \gg E_{{\rm C%
}}$, whereas for stronger tunneling, $\tilde \alpha_{{\rm t}} \gtrsim 1$,
phase fluctuations are substantially suppressed. The results derived in this
approach are valid, provided max$\{T,eV\} \gg \tilde\alpha_{{\rm t}} E_{%
{\rm C}}\exp (-2\tilde\alpha_{{\rm t}})$. This range expands rapidly with
increasing $\tilde \alpha_{{\rm t}}$. 

In conclusion, we found an effective action description of a single-electron
transistor. We analyzed it in two limits. The charge representation, which
is valid as long as $\alpha_{{\rm t}} \lesssim 1$, provides the basis for a
systematic diagrammatic description of coherent tunneling processes
including resonant tunneling. The phase representation is suitable at large
values of $\tilde\alpha_{{\rm t}} \gtrsim 1$. In both cases we calculated
the gate-voltage and temperature-dependent conductance of a single electron
transistor. The dimensionless parameters in the two limits differ by a
factor $\pi^2 \alpha_{{\rm t}} = \tilde\alpha_{{\rm t}}$. As a result the
range of validity of the two approaches overlaps and, at least
qualitatively, the two approaches cover the whole range of parameters.

The authors are grateful to D. Esteve, G. Falci and G.T. Zimanyi for useful
discussions. We thank the members of the Saclay group for sending us
their data prior to publication. 
The project was supported by the DFG within the research program of the
Sonderforschungbereich 195 and by INTAS-RFBR Grant No. 95-1305.

\begin{figure}
\centerline{\psfig{figure=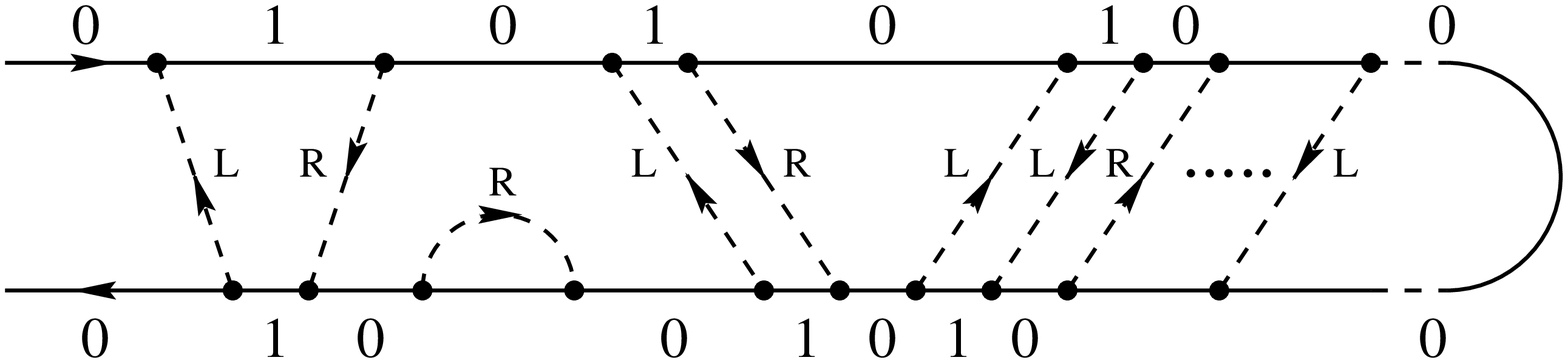,height=2.7cm}}
\caption{A diagram showing from left to right: sequential tunneling in the 
	left and right junction, a term preserving the norm, a cotunneling 
	process, and resonant tunneling.}
\label{fig1}
\end{figure}
\begin{figure}
\centerline{\psfig{figure=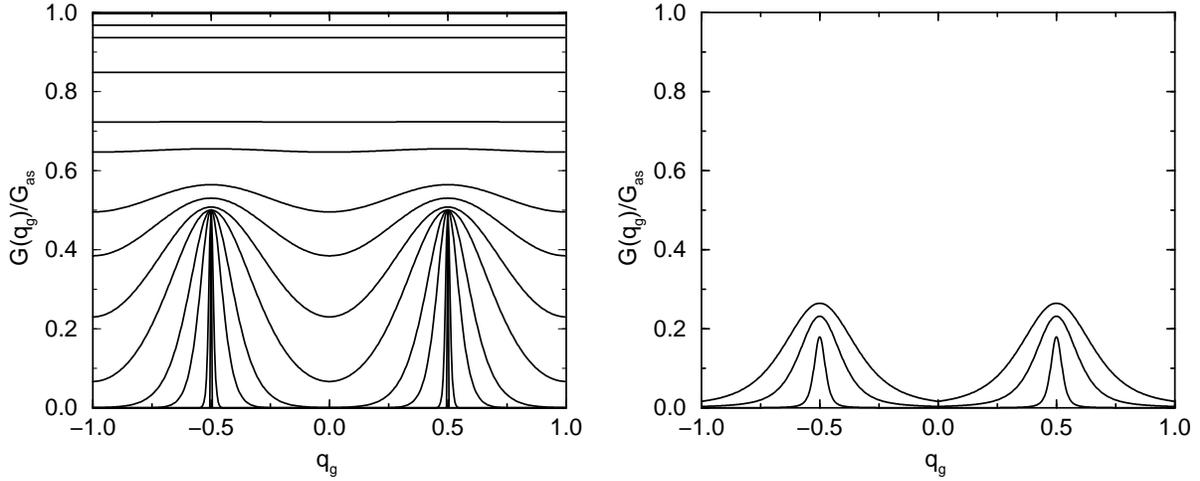,height=7.1cm}}
\caption{The linear differential conductance normalized to the high temperature
	limit.
	Left hand side: result from a master equation in lowest order 
	perturbation theory with $T/E_{\rm C} = 0.01, 0.05, 0.1, 0.2, 0.3, 0.4,
	0.5, 0.75, 1, 2, 5$, and $10$.
	In this limit the scaled conductance is independent of 
	$\alpha_{\rm t}$.
	Right hand side: result of resonant tunneling with $\alpha_{\rm t}=0.2$
	and $T/E_{\rm C} = 0.01, 0.05$, and $0.1$.
}
\label{fig2}
\end{figure}
\begin{figure}
\centerline{\psfig{figure=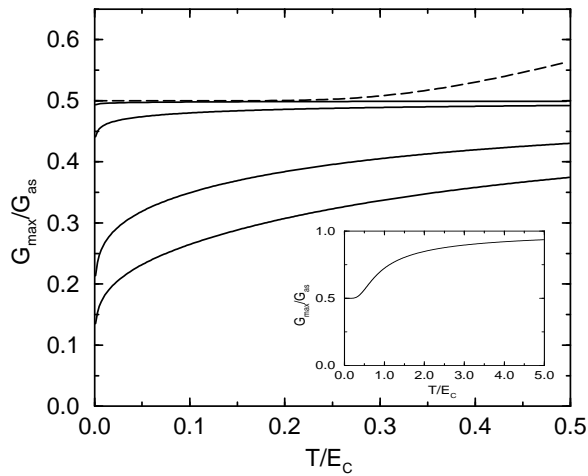,height=7.1cm}}
\caption{The maximum linear differential conductance normalized to the high 
	temperature limit for $\alpha_{\rm t}=0.001, 0.01, 0.1, 0.2$ (from top
	to bottom).
	For comparison we also show the result obtained from lowest order 
	perturbation theory (dashed line and inset).
}
\label{fig3}
\end{figure}
\begin{figure}
\centerline{\psfig{figure=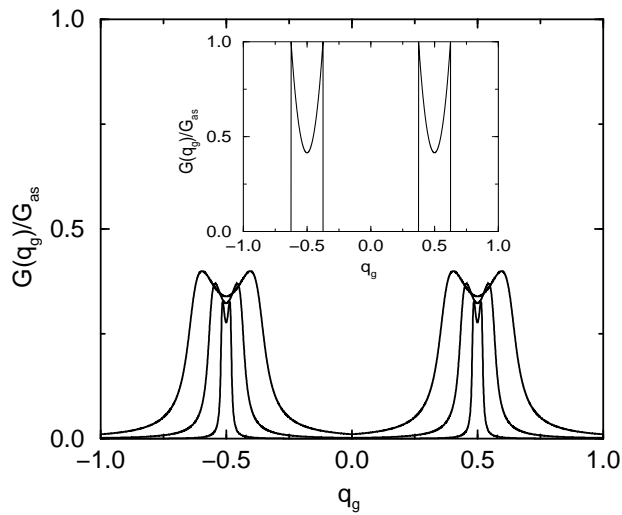,height=7.5cm}}
\caption{The normalized nonlinear differential conductance for 
	$\alpha_{\rm t}=0.1$ and $eV/E_{\rm C} = 0.05, 0.2, 0.5$ at zero
	temperature.
	The inset shows the result from a master equation in lowest order 
	perturbation theory for $eV/E_{\rm C} = 0.5$.
}
\label{fig4}
\end{figure}
\begin{figure}
\centerline{\psfig{figure=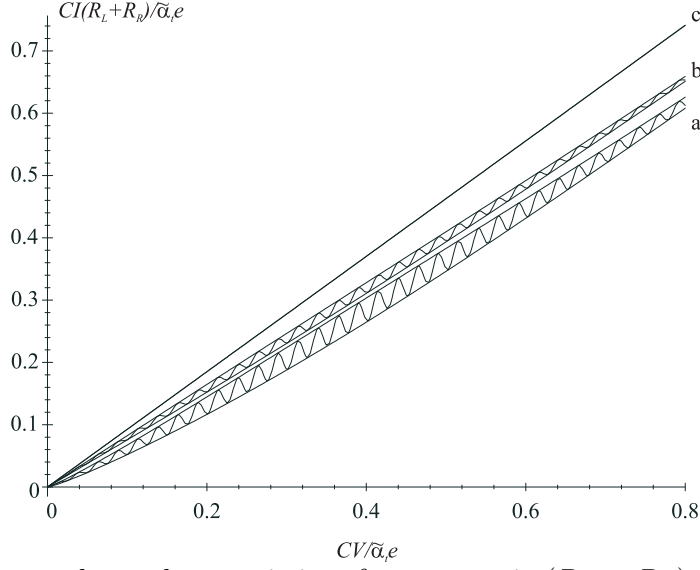,height=7.5cm}}
\caption{The current-voltage characteristics of a symmetric 
($R_{\rm L}=R_{\rm R}$) SET transistor in the strong tunneling
regime defined by eq. (\ref{vax3}) at $\tilde \alpha_{\rm t}=2.5$ 
and different
temperatures $T/ \tilde \alpha_{\rm t}E_{\rm C}=$ 0.01 (a), 0.2 (b), 
and 1 (c). 
For each temperature maximum and minimum currents are plotted. Curves with 
oscillations demonstrate the gate modulation effect. This effect is
completely suppressed in the high temperature limit (curve (c)).
}
\label{fig5}
\end{figure}
\begin{figure}
\centerline{\psfig{figure=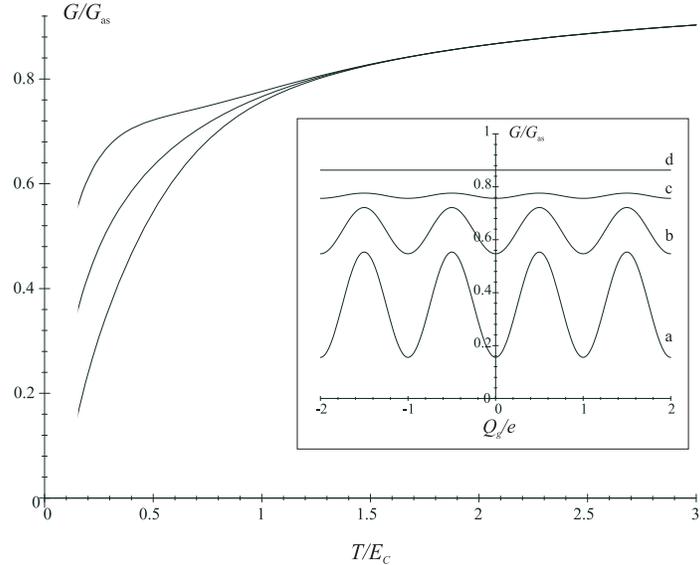,height=7.5cm}}
\caption{Maximum and minimum linear conductance of a SET transistor as a 
function of temperature obtained from the Langevin equation analysis 
(eq. (\ref{G})) for $\tilde \alpha_{\rm t}=2$. The intermediate curve
shows the linear conductance averaged over all values of the gate charge.
Inset: Conductance as a function of the gate charge for the same 
$\tilde \alpha_{\rm t}$ at different temperatures 
$T/ E_{\rm C}=$ 0.15 (a), 0.5 (b), 1 (c) and 2 (d).}
\label{fig6}
\end{figure}
\begin{figure}
\centerline{\psfig{figure=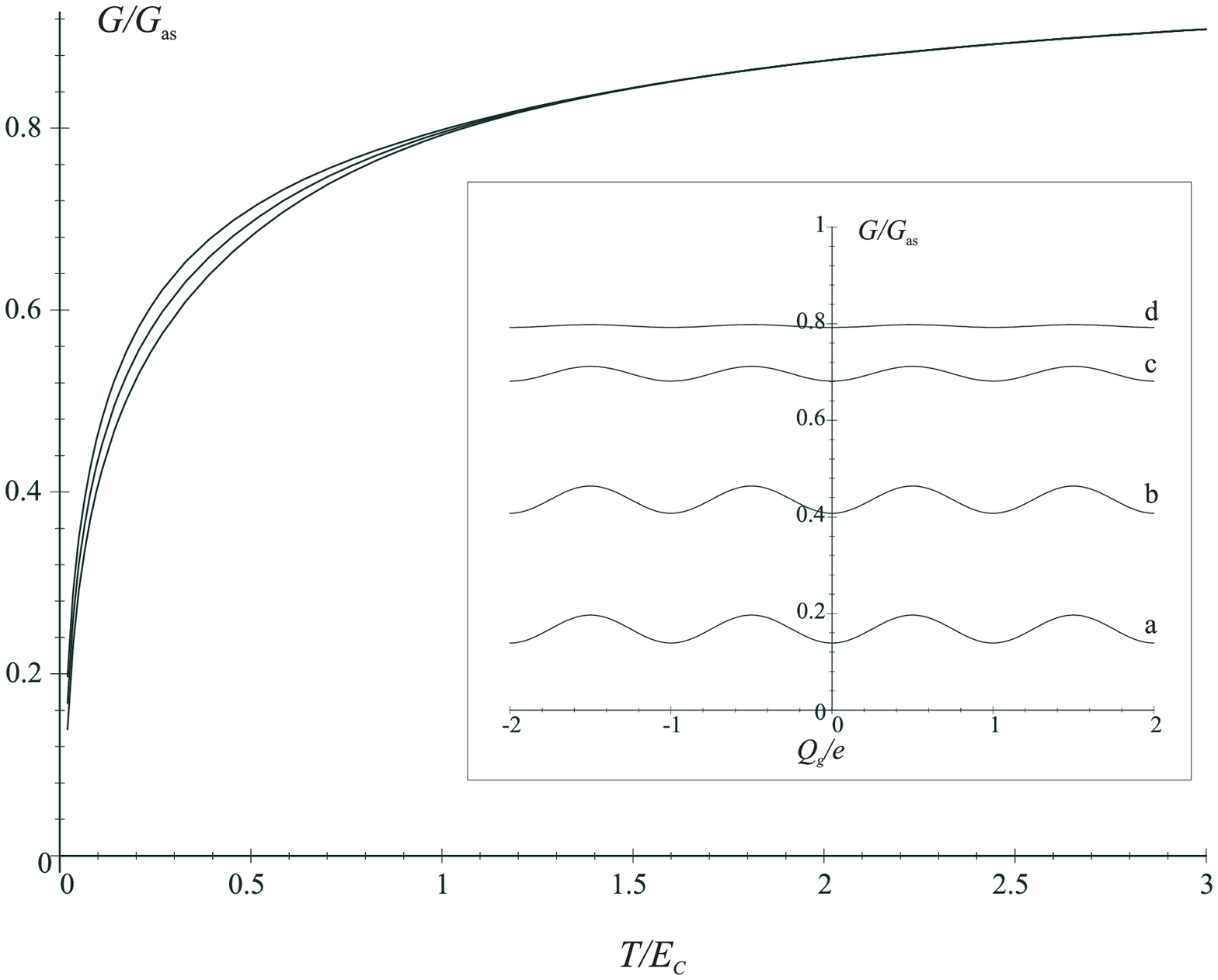,height=7.5cm}}
\caption{The same as in fig. 6 for $\tilde \alpha_{\rm t}=3$. The curves 
(a), (b), (c) and (d) in the inset correspond respectively to 
$T/ E_{\rm C}=$ 0.02, 0.1, 0.5 and 1.}
\label{fig7}
\end{figure}

\pagebreak


\begin{references}
\bibitem{Fn}  In Ref.~\onlinecite{PZ91} the function $\beta $ 
was derived for a
more general case of a tunnel junction interacting with a linear Ohmic
dissipative environment.
\end{references}
\end{document}